\documentclass[aps,preprintnumbers,amsmath,amssymb,prl,superscriptaddress,longbibliography,twocolumn]{revtex4-1}
\usepackage{dcolumn}
\usepackage{color}
\usepackage[caption=false]{subfig}
\usepackage{feynmp}
\usepackage{feynmp-auto}
\usepackage{systeme,mathtools}
\usepackage{easyReview}
\usepackage{physics}
\usepackage{dsfont}

\def\comment#1{}

\newcommand{\nc}{\newcommand}
\nc{\beq}{\begin{eqnarray}}
	\nc{\eeq}{\end{eqnarray}}
\nc{\scs}{\scriptstyle}
\nc{\setval}{\fmfset{wiggly_len}{3mm} \fmfset{arrow_len}{1.5mm}
	\fmfset{arrow_ang}{13} \fmfset{dash_len}{1.5mm}\fmfpen{0.125mm}
	\fmfset{dot_size}{2thick}}

\usepackage{bm,latexsym,mathrsfs,enumerate,color}
\usepackage[mathcal]{euscript}
\usepackage[breaklinks=true,unicode=true,urlcolor = blue,colorlinks = true,citecolor = blue,linkcolor = blue]{hyperref}
\usepackage{graphicx}
\usepackage{todonotes}
\usepackage{wrapfig}
\usepackage{easyReview}
\usepackage{cleveref}
\usepackage{tikz}
\usetikzlibrary{snakes}
\usepackage{nicefrac}

\renewcommand{\vec}[1]{\bm{#1}}

\def\slashchar#1{\setbox0=\hbox{$#1$}           
	\dimen0=\wd0                                 
	\setbox1=\hbox{/} \dimen1=\wd1               
	\ifdim\dimen0>\dimen1                        
	\rlap{\hbox to \dimen0{\hfil/\hfil}}      
	#1                                        
	\else                                        
	\rlap{\hbox to \dimen1{\hfil$#1$\hfil}}   
	/                                         
	\fi}                       %

\DeclareMathAlphabet\mathbfcal{OMS}{cmsy}{b}{n}

\renewcommand{\i}{\text{i}}

\renewcommand{\d}[2][]{\text{d}^{#1}#2}

\newcommand{\Cal}[1]{{\cal #1}}
\newcommand{\1}{\mathds{1}}

\begin{document}

	\title{Geometric phases distinguish entangled states in wormhole quantum mechanics}

	\author{Flavio S. Nogueira}
	
	\affiliation{Institute for Theoretical Solid State Physics, IFW Dresden, 01069 Dresden, Germany}
	
	\author{Souvik Banerjee}
	
	\affiliation{Institute for Theoretical Physics and Astrophysics,
		Julius-Maximilians-Universit\"at W\"urzburg, 97074 W\"urzburg, Germany}
	
	\affiliation{W\"urzburg-Dresden Cluster of Excellence ct.qmat}
	
	\author{Moritz Dorband}
	
	\affiliation{Institute for Theoretical Physics and Astrophysics,
		Julius-Maximilians-Universit\"at W\"urzburg, 97074 W\"urzburg, Germany}
	
	\affiliation{W\"urzburg-Dresden Cluster of Excellence ct.qmat}
	
	\author{\\Ren\'e Meyer}
	
	\affiliation{Institute for Theoretical Physics and Astrophysics,
		Julius-Maximilians-Universit\"at W\"urzburg, 97074 W\"urzburg, Germany}
	
	\affiliation{W\"urzburg-Dresden Cluster of Excellence ct.qmat}
	
	\author{Jeroen~van~den~Brink}
	
	\affiliation{Institute for Theoretical Solid State Physics, IFW Dresden, 01069 Dresden, Germany}
	
	\affiliation{W\"urzburg-Dresden Cluster of Excellence ct.qmat}
	
	\affiliation{Institute for Theoretical Physics, TU Dresden, 01069 Dresden, Germany}
	
	\author{Johanna Erdmenger}
	
	\affiliation{Institute for Theoretical Physics and Astrophysics,
		Julius-Maximilians-Universit\"at W\"urzburg, 97074 W\"urzburg, Germany}
	
	\affiliation{W\"urzburg-Dresden Cluster of Excellence ct.qmat}
	
	\begin{abstract}
		We establish a relation between entanglement in simple quantum mechanical qubit systems and in wormhole physics as considered in the context of the AdS/CFT correspondence. We show that in both cases, states with the same entanglement structure, indistinguishable by any local measurement, nevertheless are characterized by a different Berry phase. This feature is experimentally accessible in coupled qubit systems where states with different Berry phase are related by unitary transformations. In the wormhole case, these transformations are identified with a time evolution of one of the two throats.\end{abstract}

	\maketitle

\label{sec:intro}
In recent years, significant progress has been achieved in
establishing new relations between geometry and gravity on the one
hand, and quantum entanglement on the other. An important example is
the Ryu-Takayanagi formula \cite{Ryu:2006bv} in the context of the AdS/CFT correspondence \cite{Maldacena:1997re} 
relating the
entanglement entropy of a conformal field theory (CFT) to the area of a
minimal surface in Anti-de Sitter (AdS) space. Moreover, in the ER=EPR conjecture \cite{Maldacena:2013xja} it is argued that the entanglement in a thermofield double (TFD) state is holographically realized by a geodesic in a non-traversable wormhole in AdS space. The length of the geodesic, stretching between the two boundaries of AdS space, quantifies the amount of entanglement \cite{VanRaamsdonk:2010pw}.

Within a simpler setting, 
an early example is provided by the semiclassical Wheeler wormhole 
\cite{Wheeler_PhysRev.97.511,GARFINKLE1991146}. 
An important feature of this solution is that the magnetic field involved cannot be globally  written in terms of a vector potential. This amounts to a non-exact symplectic form, yielding a quantized flux, similarly to a magnetic monopole \cite{novikov1982hamiltonian}.

Recently, H. Verlinde \cite{Verlinde:2021kgt} studied quantum-mechanical examples of wormholes by analyzing their partition functions. 
For systems with a non-exact symplectic form, the thermal partition function becomes a functional integral over a two-dimensional surface, and corresponds to the Renyi entropy of a thermal mixed state.
\begin{figure}[t]
	\begin{center}
		\includegraphics[scale=0.7]{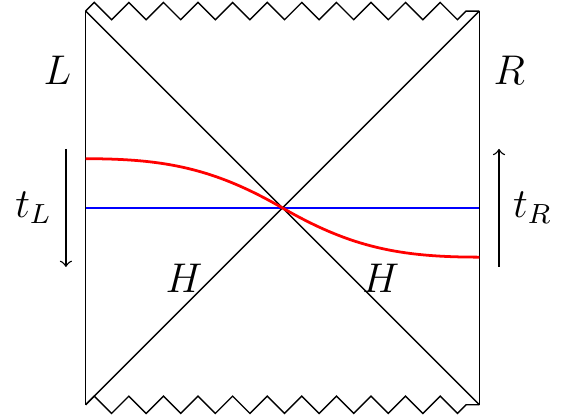}
	\end{center}
	\vspace{-0.5cm}
	\caption{Kruskal diagram of the eternal Schwarzschild black hole in AdS with time in the vertical and spatial coordinates in the horizontal direction. Vertical lines denote the left ($L$) and right ($R$) asymptotic boundaries where the CFTs are defined. The jagged lines are the singularities and diagonal lines represent the black hole horizon ($H$). The blue line corresponds to the wormhole dual to the original TFD state while the red one corresponds to a time-shifted wormhole. Arrows indicate the directions of time in the boundary theories.}
	\vspace{-0.7cm}
	\label{fig:BHandTFD}
\end{figure}

	Here  
	we show further novel similarities between a simple coupled two-qubit system in quantum mechanics and spacetime wormholes in gravity, in form of a non-factorized structure of the respective Hilbert spaces. 
	Although in quantum mechanics there is no notion of a horizon causally separating two regions, the non-factorized structure demonstrates how entanglement between the two subsystems plays the pivotal role in building the full Hilbert space. For quantum mechanics as well as for gravity, the non-factorization manifests as a class of states with the same entanglement but different Berry phases.

For the gravity analysis, we focus on the entanglement structure of an eternal black hole in AdS space. This black hole is dual to a maximally entangled state of two decoupled CFTs living on the left ($L$) and right ($R$) boundaries in the TFD state \cite{Maldacena:2001kr} (Fig. \ref{fig:BHandTFD}). This state can be derived from the Hartle-Hawking wave functional for the bulk geometry \cite{Hartle:1976tp, Israel:1976ur,Harlow:2014yka}. As shown in \cite{Papadodimas:2015xma, Verlinde:2020upt}, there exists a class of time-shifted wormholes dual to states related to the original TFD state by phases $\alpha_n$. These arise from unequal time evolution at the boundaries. From the perspective of the bulk AdS geometry, these phase-shifted states correspond to the same wormhole geometry, but with different asymptotic identifications of 
boundary times. Such phase-shifted states have the same entanglement properties as the TFD state. However, their Berry phases are different. This due to the fact that a gravitational system in the presence of black holes does not have a globally defined time. 

Here we show that by distinguishing between states with the same entanglement, the Berry phases with respect to this phase-shifted parameter provide a precise analogue to the entanglement structure of coupled quantum spins in a magnetic field. By determining the Berry phase for entangled states of two coupled spins with respect to the magnetic field, we indeed directly find that different states related by unitary transformations, and thus sharing the same entanglement structure, have nevertheless different Berry phases. Furthermore, as we will show following \cite{Verlinde:2021kgt}, in both our quantum spin system and in gravity, this behavior traces back to the presence of a non-exact symplectic form.
The quantum spin model we consider is realized for instance by a hydrogen atom with hyperfine coupling between proton and electron spin. In the quantum mechanics context, finding different Berry phases for states with the same entanglement can be probed experimentally by quantum-state tomography \cite{Q-Tomography-Review} on this specific qubit pair.  

{\it Unitary transformations and Berry phase: coupled quantum spins ---}
We first consider the entanglement structure of a hydrogen atom with hyperfine coupling $J$ between proton and electron spins in magnetic field. In first approximation, the Zeeman coupling to the proton spin can be neglected,  
leading to the Hamiltonian,
\begin{align}
	H=J\vec{S}_1\cdot\vec{S}_2-2\mu_BBS_{1z},\label{eq:HSpinCoupling}
\end{align}
where the second term is the electronic Zeeman interaction. The ground state with energy $E_0=-J/4-\sqrt{J^2/4+(\mu_BB)^2}$ is given by $\ket{\psi_0}=-\sin(\alpha/2)\ket{1,0}+\cos(\alpha/2)\ket{0,0}$,
in terms of the singlet and triplet states $\ket{s,m}$ and $\tan\alpha=2\mu_BB/J$. For our analysis we generalize the Hamiltonian \eqref{eq:HSpinCoupling} to an arbitrary coupling term $\propto\vec{B}\cdot\vec{S}_1$. Then we apply unitary transformations on the spins designed such that the interaction term reduces back to the form in \eqref{eq:HSpinCoupling}. While this restricts the transformation on the first spin to
	$U_1(\varphi,\theta)=e^{-\i\varphi S_{1z}}e^{-\i\theta S_{1y}}e^{\i\varphi S_{1z}}$,
we are free to choose what transformation to use on the second spin. Thus, we set  $U=U_1\otimes U_2$, where  $U_2(\varphi,\theta)=U_1(\lambda\varphi,\lambda\theta)$, with a parameter $\lambda\in[0,1]$ interpolating between $U_2=\1$ for $\lambda=0$ and $U_2=U_1$ for $\lambda=1$.

The entanglement entropy between the spins is not affected by the application of a local unitary transformation $U=U_1\otimes U_2$, in which $U_1$ and $U_2$ only act on their respective subsystem \cite{bengtsson2017geometry}, i.e., 
		$S_{\text{EE}}^i\big(\tr_j(\ket{\psi_0}\bra{\psi_0})\big)=S_{\text{EE}}^i\big(\tr_j(U\ket{\psi_0}\bra{\psi_0}U^\dagger)\big)$,
	where $i,j\in\{1,2\}$ and $i\neq j$. 
	Thus, 
	we obtain a class of states, parametrized by $\lambda$, with the same entanglement entropy. As we now discuss, these states are nevertheless distinguished by the Berry phase, which is sensitive to $\lambda$.

{\it Berry phase probing the hyperfine structure of entanglement ---}	
To define the Berry connection for the setup discussed above, we make use of the Maurer-Cartan form \cite{Nakahara:2003nw}. This is the natural connection on a group manifold $\Cal{M}$ and is defined for any group element $\sigma\in\Cal{M}$ as $A_{\text{MC}}=\sigma^{-1}\d{\sigma}$. The group element we choose in our example is the unitary transformation $U$ and d is the exterior derivative acting on the phase space variables $\varphi$ and $\theta$. The Berry connection is then given by the expectation value of the Maurer-Cartan form for the ground state, 
\begin{align}
	A_{\text{B}}(\lambda)&=\i\bra{\psi_0}A_{\text{MC}}\ket{\psi_0}=\i\bra{\psi_0}\left(U^\dagger\d{U}\right)\ket{\psi_0}\notag\\
	&=\frac{\sin\alpha}{2}\left\{(1-\cos\theta)-\lambda\left[1-\cos(\lambda\theta)\right]\right\}\d{\varphi}.\label{eq:BerryConnection}
\end{align}
From it
we define the associated symplectic (Kirillov-Kostant) form, 
$\omega_{\text{KK}}=\d{A}_{\text{MC}}$ \cite{Nakahara:2003nw}. 
Hence, 
\begin{align}
	F_{\text{B}}(\lambda)&=\i\bra{\psi_0}\omega_{\text{KK}}\ket{\psi_0}\notag\\
	&=\frac{\sin\alpha}{2}\left(\sin\theta-\lambda^2\sin(\lambda\theta)\right)\d{\theta}\wedge\d{\varphi}.\label{eq:BerryCurvature}
\end{align}
The Berry phase follows by integrating \eqref{eq:BerryCurvature} over the phase space
\begin{align}
	\Phi_{\text{B}}=\hspace{-0.2em}\int_0^{2\pi}\hspace{-0.3em}\int_0^\pi F_{\text{B}}(\lambda)=\pi\sin\alpha\{2-\lambda[1-\cos(\lambda\pi)]\},\label{eq:BerryPhase}
\end{align}
which is nontrivial as long as $\lambda\neq1$, i.e. $U_2\neq U_1$ yields a nontrivial Berry phase.
{\it The key observation is that according to \eqref{eq:BerryPhase}, two states $\ket{\psi_0}$ and $U\ket{\psi_0}$ with the same entanglement entropy 
	nevertheless have a different Berry phase, characterized by a different value of $\lambda$. This feature can be converted into an experimental tool to differentiate between two such states with indistinguishable entanglement properties.}

The nonzero Berry phase \eqref{eq:BerryPhase} is related to the non-exactness of the Berry curvature \eqref{eq:BerryCurvature}. 
The space in our case being just compact $S^2$, the symplectic form, and equivalently the Berry curvature cannot be globally exact. A nontrivial Berry phase can be understood as the holonomy \cite{PhysRevLett.51.2167} associated to this non-exact symplectic form.

{\it Berry phase and non-exact symplectic structure ---}
Let us compare the above results to gravitational wormholes, focussing on the non-factorization of spacetime in the presence of a wormhole. To do so, we reformulate the symplectic structure of the Hamiltonian \eqref{eq:HSpinCoupling} and the unitary transformation $U$ by a group theoretic analysis.

A single spin 
is represented in $\mathbf{CP}^1$ variables $z_i$ as,
	$\mathbf{S}=(1/2)z_i^\ast\boldsymbol{\sigma}^{ij}z_j$.
For the two-spin model 
we may introduce $\mathbf{CP}^1$ variables for each spin, which in absence of an external field leads to a tensor product structure $\mathbf{CP}^1\times\mathbf{CP}^1$ that can 
be realized as a diagonal embedding into $\mathbf{CP}^3$. The factorized $\mathbf{CP}^1\times\mathbf{CP}^1$ structure is sufficient to explain local properties.  
	However to understand non-local correlations, leading to wormhole-like structures, it is necessary to consider the full $\mathbf{CP}^3$ symmetry. In particular, including the Zeeman term as in \eqref{eq:HSpinCoupling} breaks the tensor product structure. Using the $\mathbf{CP}^3$ representation we can understand the Berry phase arising for different unitaries.

As a coset space $\mathbf{CP}^3=\text{SU}(4)/\text{U}(3)$, so the U(1) factor in $\text{U}(3)$ can be used to define a U(1) bundle. The bundle connection follows from the Maurer-Cartan form, with symplectic form given by the exterior derivative of the connection. Since in the presence of the magnetic field, the ground state 
does not lie inside the natural $\mathbf{CP}^1\times\mathbf{CP}^1\subset\mathbf{CP}^3$ submanifold, we 
have  
to consider the Berry connection inside the full U(1) bundle given by $\mathbf{CP}^3$. In particular, the state evolution given by $U(\lambda)=U_1\otimes U_2(\lambda)$ 
implies a Berry connection \eqref{eq:BerryCurvature} on the full $\mathbf{CP}^3$ bundle.

For a general 
$g\in\text{SU}(4)$ 
in the coset construction of $\mathbf{CP}^3$, the connection in the U(1) fibre is defined as
\begin{align}
	A=\i\tr(t_{\text{U}(1)}A_{\text{MC}}(g)).\label{eq:CP3Connection}
\end{align}
Here, $t_{\text{U}(1)}$ corresponds to the generator of SU(4) yielding the U(1) factor of U(3) given by
\begin{align}
	t_{\text{U}(1)}=\frac{1}{\sqrt{24}}\text{diag}(1,1,1,-3).\label{eq:U1Factor}
\end{align}
To obtain the connection for the Hamiltonian \eqref{eq:HSpinCoupling}, we pick the transformation $U$ as group element of SU(4). 
This group element  
only belongs to the factorised subgroup $\text{SU}(2)\times\text{SU}(2)$. However, SU(4) $\supset$ SU(2) $\times$ SU(2) $\times$ U(1) where the U(1) factor rotates in the first SU(2) subspace with a positive phase, and in the second SU(2) subspace with the opposite phase \cite{georgi1999lie}. To account for this relative sign in our transformation $U(\lambda)$, we have to replace $\sigma_i\to-\sigma_i$ in the transformation acting only on the second spin, $U_2\to\tilde{U}_2= U_2{}^T$. Inserting \eqref{eq:U1Factor} into \eqref{eq:CP3Connection} and considering $g=U_1\otimes\tilde{U}_2$, we find the same connection as 
in \eqref{eq:BerryConnection}
\cite{SupMat}.

Following the argument of \cite{Verlinde:2021kgt}, we identify the non-exact Berry curvature \eqref{eq:BerryCurvature} as 
signaling  
a wormhole structure. As discussed above, this structure arises from non-diagonally embedded submanifolds of $\mathbf{CP}^3$. It is not manifest for the diagonal embedding $\mathbf{CP}^1\times\mathbf{CP}^1$.

{\color{black}
	We now show that the Berry phase non-exact symplectic structure 
	is also present in actual spacetime wormholes due to the lack of a global time-like Killing vector. 
	Therefore, 
	correlations present in the emergent wormhole show up in the Berry connection whenever different unitary transformations act upon the two spins.
}

{\it Unitary transformations and Berry phase: gravitational wormholes ---}
The group-theory and entanglement structures obtained for the two-spin quantum mechanical model arise also for gravitational wormholes in an exactly analogous way. As an example, we consider an eternal black hole in AdS spacetime depicted in Fig. \ref{fig:BHandTFD}.

We find, in exact analogy to the two-spin example, a Berry phase by considering different unitary transformations given by Hamiltonian evolution of the boundary states. 
The class of wormholes shown in Fig. \ref{fig:BHandTFD} is dual to the class of states
\begin{align}
	\ket{\text{TFD}_\alpha}=\frac{1}{\sqrt{Z}}\sum_ne^{\i\alpha_n}e^{-\beta\frac{E_n}{2}}\ket{n}_L\ket{n}_R^\ast,\label{eq:TFD}
\end{align}
where $Z$ is the partition sum, $\beta$ is the inverse temperature and $E_n=(E_n^{(L)}+E_n^{(R)})/2$ are the sums of the energy eigenvalues corresponding to the left and right energy eigenstates $\ket{n}_{L/R}^{(\ast)}$. When all phases $\alpha_n$ vanish, 
\eqref{eq:TFD} reduces to the standard TFD state. These additional phase factors can be incorporated in the Hartle-Hawking derivation when relating left and right boundary states $\ket{n}_L$ and $\ket{n}_R$ by an anti-unitary operator $\Theta$ \cite{Hartle:1976tp}. The map $\ket{n}_L=\Theta\ket{n}_R$ is not unique. The fact that it is defined up to a phase gives rise to the phases in \eqref{eq:TFD} \cite{SupMat},  
which do not change the entanglement properties of the resulting state. When calculating the reduced density matrix of either CFT using \eqref{eq:TFD}, the phases $\alpha_n$ drop out,  
causing  
the entanglement entropy 
to remain insensitive 
\cite{SupMat}. 
The same conclusion is also reached by computing the mixed two-point correlators $\bra{{\text{TFD}_\alpha}}\Cal{O}_L\Cal{O}_R \ket{\text{TFD}_\alpha}$. They  
are insensitive to the phases $\alpha_n$ \cite{Papadodimas:2015jra}, implying unaltered entanglement properties.

Each field theory boundary shown in Fig. \ref{fig:BHandTFD}
\begin{figure}[t]
	\begin{center}
		\includegraphics[scale=0.7]{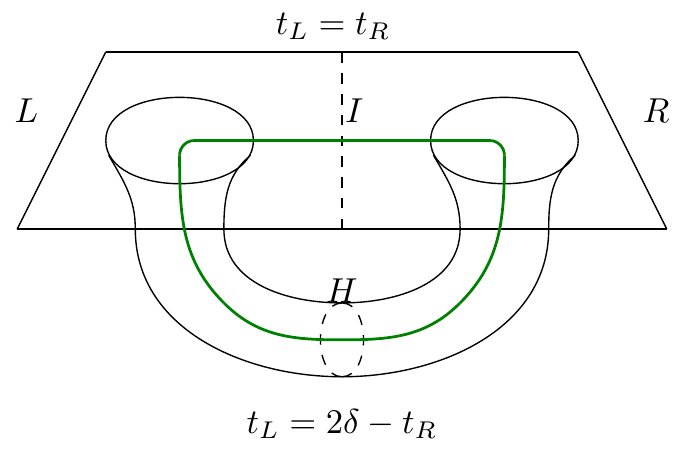}
	\end{center}
	\vspace{-0.5cm}
	\caption{Schematic representation of a wormhole corresponding to the colored lines in Fig. \ref{fig:BHandTFD}. The times $t_L$ and $t_R$ of the left and right regions are identified at the interface $I$. In the bulk a different relation is used that accounts for the sign flip of the time-like Killing vector at the horizon $H$. The Berry phase is accumulated along the closed loop printed in green.}
	\vspace{-0.6cm}
	\label{fig:Gluing}
\end{figure}
has a time coordinate $t_L$ and $t_R$, respectively. As shown in Fig. \ref{fig:Gluing}, at the interface $I$ we have to specify an identification between the boundary times, which for convenience we choose as $t_L=t_R$. However, since there is no globally defined time for the dual geometry due to the presence of the black hole horizon $H$, this asymptotic identification cannot hold throughout the
bulk. In particular, since the timelike Killing vector flips sign across the horizon, a time shift $\delta$ at the horizon needs to be taken into account, via the relation $t_L=2\delta-t_R$ \cite{Verlinde:2020upt}. Together with the boundary identification $t_L=t_R$, this relation implies $\delta=t_L=t_R$ at the boundaries. Thus, time evolution on both boundaries can be expressed using the emerging parameter $\delta$. As shown in \cite{Verlinde:2020upt}, the additional phases appearing in the time-shifted TFD states \eqref{eq:TFD} can be understood as resulting from time evolution through an identification $\alpha_n=-2E_n\delta$. In \eqref{eq:TFD}, the $\alpha_n$ clearly have periodicities of $2\pi$,
\begin{align}
	\ket{\text{TFD}_\alpha}=\ket{\text{TFD}_{\alpha+2\pi}}.\label{eq:Periodicity}
\end{align}
Thus, in analogy to the quantum mechanics example, we define a Berry connection for \eqref{eq:TFD}, with $\alpha_n=-2E_n\delta$,
\begin{align}
	A_\delta=\i\bra{\text{TFD}_\alpha}\partial_\delta\ket{\text{TFD}_\alpha}=\frac{2}{Z}\sum_nE_ne^{-\beta E_n}.\label{eq:BerryConnectionTFD}
\end{align}
\vspace{-0.5pt}Note that $\delta$ is a phase space variable. We stress here that the existence of $\delta$ is a direct consequence of the lack of a global time in the gravitational spacetime.

Eq. \eqref{eq:BerryConnectionTFD} follows from considering unitary transformations in the asymptotic symmetry group, which is the subgroup of bulk diffeomorphisms leaving the boundary conditions invariant. The gauge parameter of this group is $\delta$  
\cite{SupMat}.

In order to show the analogy between the 
two-spin system 
and the gravitational example, we consider below two contrasting scenarios. 

({\it i}) $\lambda=0$: In this case the  
quantum mechanical example leading to \eqref{eq:BerryConnection}  
implies a nontrivial Berry connection.  
An analogous situation arises for the TFD state for the unitary operation
\begin{align}
	u_0(\delta)=e^{-\i(H_L+H_R)\delta}\label{eq:DifferentTrafo}
\end{align}
belonging to the asymptotic symmetry group. The transformation \eqref{eq:DifferentTrafo} can be turned into a one-sided transformation 
via
\begin{align}
	\tilde{u}_0(\delta)&=u_1(\delta)\cdot u_0(\delta)=e^{-2\i H_L\delta}\label{eq:OneSidedTrafo},\\
	u_1(\delta)&=e^{-\i(H_L-H_R)\delta},\label{eq:SameTrafo}
\end{align}
where $u_1(\delta)$ is a trivial transformation for $H_L=H_R$.
In order to compute the Berry connection,  
we note that the time-shifted TFD states \eqref{eq:TFD} are obtained from the TFD states without additional phases
by applying \eqref{eq:OneSidedTrafo},
\begin{align}
	\ket{\text{TFD}_\alpha}=\tilde{u}_0(\delta)\ket{\text{TFD}_{\alpha=0}}.
\end{align}
With respect to 
$\delta$, the Berry connection is then given by \eqref{eq:BerryConnectionTFD}. This is the exact analogue of the quantum mechanical case with $\lambda=0$. Both for the two-spin system and for the wormhole geometry,  
nontrivial Berry connections are obtained for one-sided transformation, which can also be understood as a nonzero holonomy \footnote{In differential geometry, a non-trivial holonomy arises when the considered manifold has to be described by more than one coordinate patch, each of which patch has its own connection. Therefore, one cannot define a symplectic form as an exterior derivative of a single connection. Hence, the symplectic form cannot be globally exact. However, the connections on individual patches are related to each other by U(1) gauge transformations and yields a nonzero Berry phase. In the case of gravity with a horizon, the identifications $\delta=t_L$ and $\delta=t_R$ only work at the left and right boundaries and can at most be analytically continued in the near-boundary regions, leading to two different coordinate patches. The connections on the coordinate patches are related by a unitary of the same form as \eqref{eq:DifferentTrafo}. This can be interpreted as a holonomy of a U(1) bundle, as in the quantum mechanical case, and results in a non-vanishing Berry phase \cite{PhysRevLett.51.2167}. For the TFD state, this non-vanishing Berry connection can also be argued in terms of the holonomy associated to the modular Hamiltonian \cite{Czech:2017zfq,Czech:2018kvg,Czech:2019vih}}.

({\it ii}) $\lambda=1$: According to \eqref{eq:BerryConnection}, the $\lambda=1$ case in the quantum mechanics example 
implies 
a vanishing Berry connection. An analogous situation arises for the TFD state when we consider the unitary operation \eqref{eq:SameTrafo}. Since the times $t_L$ and $t_R$ run in opposite direction \footnote{This follows from the fact that the bulk Killing vector switches sign at the horizon. This is accounted for by relating the CFT times at the left and right boundaries to the Schwarzschild times in the respective bulk wedges with a relative flip of sign. Note that this is independent of any additional phases in the TFD state \eqref{eq:TFD}.} as shown in Fig. \ref{fig:BHandTFD}, the transformation \eqref{eq:SameTrafo} corresponds to applying the same unitary transformation to the two subsystems as in the two-spin system. 
The Berry connection for this transformation 
reads,
	$A_\delta=\i\bra{\text{TFD}_{\alpha=0}}u_1^\dagger\partial_\delta u_1\ket{\text{TFD}_{\alpha=0}}=0$,
in analogy to the $\lambda=1$ case in \eqref{eq:BerryConnection}.

{\it Non-exact symplectic form within gravity ---} These structures may be realized explicitly for a wormhole in Jackiw-Teitelboim (JT) gravity \cite{JACKIW1985343,TEITELBOIM198341,Sarosi:2017ykf}, consisting of AdS$_2$ gravity coupled to a dilaton 
field. The Hamiltonians $H_L$ and $H_R$ are given in terms of the ratio of the dilaton values at the horizon and the AdS boundary, $\phi_h$ and $\phi_b$, as $H_L=H_R=\phi_h^2/\phi_b{\color{green!40!black}~>0}$ \cite{Harlow:2018tqv}.
Thus, the asymptotic symmetries are given only by time translations with associated gauge parameter $\delta$. Clearly, the difference $H_L-H_R$ is a trivial operator and only $H=H_L+H_R$ generates time evolution. The associated phase space consists of variables $\delta$ and $\phi_h$, or equivalently $\delta$ and $H$, with symplectic form \cite{Harlow:2018tqv}
\begin{align}
	\omega=\frac{4\phi_h}{\phi_b}\d{\delta}\wedge\d{\phi_h}=\d{\delta}\wedge\d{H}.\label{eq:SympForm}
\end{align}
As in the quantum mechanical example, this symplectic form 
yields the Berry curvature. 
The trajectory for the Berry phase is visualized in
Fig. \ref{fig:Gluing}. It enters the wormhole on one side, exits on the other and then closes the loop to the starting point outside the wormhole. Recall that due to the presence of the horizon, defined by the region where the dilaton $\phi$ takes the value $\phi_h$, $\delta$ is $\nicefrac{\pi}{E}$-periodic following \eqref{eq:Periodicity} with $E=\nicefrac{\phi_h^2}{\phi_b}$, and $H>0$ \footnote{Note that the wormhole structure is supported by a non-vanishing $\phi_h$ only; setting $\phi_h=0$ collapses the geometry to pure AdS$_2$.}. 
From this 
we find that the connection \eqref{eq:BerryConnectionTFD} and its symplectic form \eqref{eq:SympForm} are defined only on the complement of the origin of $\{\delta,H\}$ phase space, i.e. the phase space has the topology of $\mathbf{R}^2\setminus\{0\}$, where $\delta$ and $\phi_h\propto H$ correspond to angular and radial coordinate,  respectively. 
In JT gravity, where $E=\nicefrac{\phi_h^2}{\phi_b}$ and \eqref{eq:BerryConnectionTFD} evaluates to $A_\delta=2\nicefrac{\phi_h^2}{\phi_b}$, this yields a Berry phase
\begin{align}
	\Phi_{\text{B}}^{\text{JT}}=\oint A_{\delta}\d{\delta}=2\frac{\phi_h^2}{\phi_b}\int_0^{\frac{\pi}{E}}\d{\delta}=2\pi,\label{eq:BerryPhaseJT}
\end{align}
associated with the holonomy of the closed path. Since $\delta$ is periodic while $\phi_h$ is not, this holonomy is understood as a winding number of a circle,  
thus establishing that the symplectic structure \eqref{eq:SympForm} is non-exact. The holonomy \eqref{eq:BerryPhaseJT} is due to the fact that in presence of the horizon, the path through the wormhole cannot be shrunk to a point. Eq. \eqref{eq:BerryPhaseJT} is analogous to the quantum mechanical result \eqref{eq:BerryPhase} for the special case 
$\lambda=0$.

{\it Discussion and conclusion ---}
The non-exactness of the symplectic structure and the corresponding Berry phases associated to a gravitational wormhole are equivalently found in analyzing the quantum mechanical two-spin system. This further exemplifies the surprising similarity between quantum mechanical models and gravitational wormholes. In the gravitational setup, wormholes arise naturally from the non-local structure of the spacetime due to the presence of the black hole horizon. In quantum mechanical systems, where locality is manifest, this feature is realized by intricate group theoretical structures. 
Using group theoretic arguments, we have actually identified the entangled degrees of freedom in our quantum mechanical system which are responsible for creating a wormhole geometry in spin space. This may be viewed as a manifestation of {\it entanglement creating spacetime} which lies at the heart of understanding AdS/CFT duality \cite{Maldacena:2013xja,VanRaamsdonk:2010pw,VanRaamsdonk:2009ar,Czech:2012bh}.

In particular, we found that both for wormholes and in simple qubit systems, there are entangled states 
having different Berry phases 
sharing the same entanglement entropy. 
This analysis might be probed on a number of experimental platforms. Generally measurements of both quantum entanglement and of Berry phases involve interference between the original starting state and a rotated one for an ensemble of identical quantum states. Apart from the two-spin systems 
discussed above, 
which are toy models for the multiple spin-qubits accessible in liquid-state nuclear magnetic resonance~\cite{Ryan2009},
also quantum dots coupled to an optical cavity~\cite{Imamoglu1999} and superconducting quantum circuits offer experimental platforms for quantum tomography on controlled qubit pairs~\cite{Steffen2006,Devoret2013,Song2017}.
A further avenue 
arises from the fact that TFD states can be experimentally prepared to a high accuracy using quantum approximate optimization algorithm for transverse field Ising models \cite{Zhu:2019bri}. Modifying this algorithm to realize time-shifted TFD states as in \eqref{eq:TFD} will provide an experimental realization of the proposed entanglement structure in this context.  

\begin{acknowledgments}
	{\it Acknowledgments ---}
	We thank Emma Loos, Björn Trauzettel and Anna-Lena Weigel for useful discussions. S.B., M.D., R.M., J.v.d.B. and J.E. acknowledge support by the Deutsche Forschungsgemeinschaft (DFG, German Research Foundation) under Germany's Excellence Strategy through the W\"urzburg-Dresden Cluster of Excellence on Complexity and Topology in Quantum Matter - ct.qmat (EXC 2147, project-id 390858490). The work of S.B., M.D., R.M. and J.E. is furthermore supported via project-id 258499086 - SFB 1170 `ToCoTronics'. The work of S.B. is supported by the Alexander von Humboldt postdoctoral fellowship.
\end{acknowledgments}

\newpage

\renewcommand{\theequation}{A.\arabic{equation}}
\setcounter{equation}{0}

\begin{center}
	\textbf{Supplemental Material}
\end{center}

{\it Bundle connection in the $\mathbf{CP}^3$ analysis ---} Here we explain in more detail the calculation of the connection in the $\mathbf{CP}^3$ analysis below eq. (5) in the main text.

The transformation $U_1$ acting on the first spin is given in eq. (3) in the main text. $U_2$ is then given by $U_2(\varphi,\theta)=U_1(\lambda\varphi,\lambda\theta)$. The U(1) factor in the decomposition of SU(4) leads to a negative sign for the second SU(2) compared to the first one \cite{georgi1999lie}. To account for this relative sign in the transformation $U_2$ acting on the second spin, we have to replace $\sigma_i\to-\sigma_i$ in $U_2$,
\begin{align}
	U_2\stackrel{\sigma\to-\sigma_i}{\to}\tilde{U}_2=e^{\i\lambda\frac{\varphi}{2}\sigma_z}e^{\i\lambda\frac{\theta}{2}\sigma_y}e^{-\i\lambda\frac{\varphi}{2}\sigma_z}=U_2{}^T.
\end{align}
We observe here that the same expression for $\tilde{U}_2$ can be obtained from $U_2$ by reversing the sign of $\lambda$. Using $\tilde{g}=U_1\otimes\tilde{U}_2$ in eq. (5) with eq (6) from the main text, we obtain, with a proper normalization factor,
\begin{align}
	A=\frac{\sin\alpha}{2}\{(1-\cos\theta)-\lambda(1-\cos(\lambda\theta))\}\d{\varphi}.
\end{align}
This is the same expression as in eq. (2) from the main text which vanishes for $\lambda=1$.

{\it Time-shifted TFD states from a path integral ---} Here we briefly review how the generalised TFD state as in eq. (10) in the main text is computed from a path integral. We discuss in more detail how the additional phase factors $\alpha_n$ which lead to the Berry phase can be naturally incorporated in this derivation.

We first sketch how the TFD state without additional phase factors (i.e. eq. (7) in the main text with $\alpha_n=0~\forall n$) arises as the groundstate $\ket{\Omega}$ at infinite past from a path integral \cite{Harlow:2014yka}. This calculation leads to the Hartle-Hawking wave function $\Psi_{\text{HH}}=\braket{\phi}{\Omega}$. Up to a normalisation factor, this wave function is defined by a Euclidean path integral
\begin{align}
	\Psi_{\text{HH}}\propto\int\Cal{D}g\Cal{D}\phi\,\exp(-S_E[g,\phi]),
\end{align}
where $\phi$ is some matter field (for instance, a real scalar) and $g$ specifies the geometry on which $\phi$ propagates. The calculation of the path integral relies on saddle point approximations for the metric. In our case, we specify $g$ to the eternal Schwarzschild AdS metric. We then consider a real scalar field on that geometry. For some generic state $\ket{\kappa}$, the groundstate is calculated as
\begin{align}
	\braket{\phi}{\Omega}=\frac{1}{\braket{\Omega}{\kappa}}\lim\limits_{t_E\to\infty}\bra{\phi}e^{-t_EH}\ket{\kappa}.
\end{align}
Up to a normalisation factor this corresponds to
\begin{align}
	\braket{\phi}{\Omega}\propto\int_{\hat{\phi}(t_E=-\infty)=0}^{\hat{\phi}(t_E=0)=\phi}\Cal{D}\hat{\phi}\,\exp(-S_E[\hat{\phi}]).\label{eq:PathIntegral}
\end{align}
$S_E$ is the Euclidean action for the scalar field on the geometry $g$ and $H$ is the corresponding Hamiltonian. Instead of integrating the lower half of the Euclidean plane by time evolution using the Hamiltonian, it is convenient to make use of the Rindler decomposition. In Rindler coordinates, the Hamiltonian is replaced by the boost operator $K_x$ (we choose a boost in $x$-direction for simplicity) and the integration is along the angular direction $\varphi$. Integrating over the half plane is then easily seen to be a rotation by $\pi$. Solving the path integral \eqref{eq:PathIntegral} requires a careful consideration of the boundary conditions: in the path integral we introduced $\phi$ as the boundary condition at $t_E=0$. However using the boost operator we find that the `initial' state is on the same time slice as the `final' state. We can fix this be splitting $\phi$ across the origin as $\phi(x<0)=\phi_L$ and $\phi(x>0)=\phi_R$, that is $\phi_R$ is the initial and $\phi_L$ is the final state. This is described by an anti-unitary operator $\Theta$ containing the time reversal operator defined in QFT. Initial ($\bra{\phi_R}$) and final ($\ket{\phi_R}$) state are related by $\bra{\phi_R}=\Theta\ket{\phi_R}$. This implies that $K_x$ may act only on the left states. The solution to the path integral \eqref{eq:PathIntegral} is then written as
\begin{align}
	\braket{\phi}{\Omega}=\braket{\phi_L\phi_R}{\Omega}=\bra{\phi_L}e^{-\pi K_L}\Theta\ket{\phi_R}.
\end{align}
Inserting a complete set of eigenstates to $K_L$ and using the anti-linearity of $\Theta$ we find
\begin{align}
	\braket{\phi}{\Omega}&=\sum_n\bra{\phi_L}e^{-\pi K_L}\ket{n_L}\bra{n_L}\Theta\ket{\phi_R}\notag\\
	&=\sum_ne^{-\pi E_n}\braket{\phi_L}{n_L}\bra{\phi_R}\Theta^\dagger\ket{n_L}\label{eq:PIResult}
\end{align}
To write $\ket{\Omega}$ manifestly as a sum over left and right states one has to impose the relation $\ket{n_R}^\ast\propto\Theta^\dagger\ket{n_L}$. As already indicated, there is a freedom in choosing this relation: while an equality leads to the commonly known expression for the TFD state, we can also choose to include a phase factor $e^{-\i\alpha_n}$. Following \eqref{eq:PIResult}, each choice for a set of $\alpha_n$ corresponds to a choice of a boundary condition for the path integral, set at the asymptotic boundary. This choice simply specifies how the times in the boundary theories are related. This freedom exists since in gravity, there is no preferred origin of time. A detailed discussion on this can be found in \cite{Papadodimas:2015jra}. As we will show in the next section, this choice does not change the entanglement structure between the left and right subsystems. Applying this to \eqref{eq:PIResult}, i.e. the TFD state, we find
\begin{align}
	\ket{\text{TFD}_\alpha}=\sum_ne^{\i\alpha_n}e^{-\pi E_n}\ket{n_L}\ket{n_R}^\ast.
\end{align}
This is the non-normalized TFD state with additional phases.

Note that the explicit calculation above was done for a Rindler observer with unit acceleration, corresponding to a temperature $\beta=2\pi$. Therefore, also inserting the normalization by the partition sum $Z$, we can rewrite the result to
\begin{align}
	\ket{\text{TFD}_\alpha}=\frac{1}{\sqrt{Z}}\sum_ne^{\i\alpha_n}e^{-\beta\frac{E_n}{2}}\ket{n_L}\ket{n_R}^\ast\label{eq:TimeEvTFD}
\end{align}
as in eq. (7) in the main text.

{\it Entanglement entropy for time shifted TFD states ---} Here we show that the additional phases in the time-shifted TFD state given in eq. (7) in the main text do not change the entanglement entropy by explicit calculation of the reduced density matrix of the left subsystem $\rho_L$ (the calculation of $\rho_R$ works analogously):
\begin{align}
	&~~~~\rho_L=\tr_R\left(\ket{\text{TFD}_\alpha}\bra{\text{TFD}_\alpha}\right)\notag\\
	&=\frac{1}{Z}\hspace{-0.1em}\sum_{m,n}e^{-\frac{\beta}{2}(E_n+E_m)}e^{\i(\alpha_n-\alpha_m)}\tr_R\left(\ket{n}_L\hspace{-0.1em}\ket{n}_R\hspace{-0.1em}\bra{m}_L\hspace{-0.1em}\bra{m}_R\right)\notag\\
	&=\frac{1}{Z}\hspace{-0.2em}\sum_{m,n,k}\hspace{-0.2em}e^{-\frac{\beta}{2}(E_n+E_m)}e^{\i(\alpha_n-\alpha_m)}\ket{n}_L\hspace{-0.1em}\bra{m}_L\hspace{-0.1em}\braket{k}{n}_R\hspace{-0.1em}\braket{m}{k}_R\notag\\
	&=\frac{1}{Z}\sum_{m,k}e^{-\frac{\beta}{2}(E_k+E_m)}e^{\i(\alpha_k-\alpha_m)}\ket{k}_L\bra{m}_L\braket{m}{k}_R\notag\\
	&=\frac{1}{Z}\sum_ke^{-\beta E_k}\ket{k}_L\bra{k}_L.
\end{align}
The reduced density matrix is independent of the phase factors. It is the same as the thermal density matrix obtained from the TFD state without additional phases $\alpha_n=0~\forall n$. Since the phases drop out, the entanglement entropy will be unchanged.

{\it Asymptotic symmetries in AdS spacetime ---} Here we give more details about asymptotic symmetries in AdS spacetimes, mentioned below eq. (8) in the main text. Especially, we explain the meaning of $\delta$, arising from the boundary condition in the path integral \eqref{eq:PathIntegral}, as emergent gauge parameter of the asymptotic symmetry group.

Since AdS is a spacetime with boundary, for a well defined variational principle it is necessary to impose boundary conditions on the dynamical fields. The variation of an action on a manifold $\Cal{M}$ with AdS$_{d+1}$ background has boundary terms
\begin{align}
	\delta S\supset\int_{\partial\Cal{M}}\d[d]{x}\sqrt{\gamma}\,T_{ij}\delta\gamma^{ij}
\end{align}
where $\gamma$ is the metric induced on the boundary. $T_{ij}$ can be defined as the boundary energy momentum tensor. $\delta\gamma^{ij}$ can be fixed such that the variation of the action vanishes, that is imposing Dirichlet boundary conditions
\begin{align}
	\gamma^{ij}=\Cal{C}^{ij},\label{eq:AsyBdCond}
\end{align}
where $\Cal{C}^{ij}$ is a tensor with constant entries.

These boundary conditions are not preserved by any arbitrary diffeomorphism. The subset of diffeomorphisms which do preserve \eqref{eq:AsyBdCond} are called asymptotic symmetries. These change \eqref{eq:AsyBdCond} only by additional constants $c^{ij}$ for each component of $\Cal{C}^{ij}$, that is the variation of $\gamma^{ij}$ still vanishes.

Asymptotic symmetries are generated by Killing vectors $\xi$ that are also an isometry of the boundary metric. This implies
\begin{align}
	\gamma_\mu^{~\alpha}\gamma_\nu^{~\beta}\nabla_{(\alpha}\xi_{\beta)}=0.\label{eq:KillingEq}
\end{align}
There is however another important distinction of asymptotic symmetries depending on the values of $c^{ij}$: there are diffeomorphisms where $c^{ij}=0$, called trivial in the following, and also such where $c^{ij}\neq0$. The latter ones will lead to a change in the boundary phase space, proportional to these constants.

Since the boundary includes the time dimension, time translations will always be a part of the asymptotic symmetries. An explicit example for a time translation with $c^{ij}\neq0$ is given by
\begin{align}
	\xi_0=e^{-\i(H_L+H_R)\delta}\label{eq:BulkDiffeo}.
\end{align}
The constants $c^{ij}$ arise as the additional phases in eq. (7) in the main text. In the JT gravity example, the asymptotic symmetries are only time translations and $c^{ij}$ has only one component. Noting that $\delta$ is not a coordinate one can check that the diffeomorphism $\xi_0$ given in \eqref{eq:BulkDiffeo} satisfies \eqref{eq:KillingEq} and describes such time translations. Acting with an asymptotic symmetry on the TFD state, the constant will appear as the additional phase containing the emerging time parameter $\delta$ as $c=\alpha_n=-2E_n\delta$ as in eq. (7) in the main text. If $\delta=0$, the TFD state does not receive any additional phase and the associated diffeomorphism is trivial.

\bibliography{main}

\end{document}